Importance of diagnostic accuracy in big data: False-positive diagnoses of type 2 diabetes in health insurance claims data of 70 million Germans

Ralph Brinks[1,2,*], Thaddäus Tönnies[2], Annika Hoyer[3]

1 Chair for Medical Biometry and Epidemiology, Faculty of Health/School of Medicine, Witten/Herdecke University, Witten, Germany

2 Institute for Biometry and Epidemiology, German Diabetes Center, Düsseldorf, Germany

3 Biostatistics and Medical Biometry, Medical School OWL, Bielefeld University, Bielefeld, Germany

Correspondence:

ralph.brinks@uni-wh.de

Article type: Original research

ABSTRACT

Large data sets comprising diagnoses about chronic conditions are becoming increasingly available for research purposes. In Germany, it is planned that aggregated claims data including medical diagnoses from the statutory health insurance with roughly 70 million insurants will be published on a regular basis. Validity of the diagnoses in such big data sets can hardly be assessed.

In case the data set comprises prevalence, incidence and mortality, it is possible to estimate the proportion of false positive diagnoses using mathematical relations from the illness-death model. We apply the method to age-specific aggregated claims data from 70 million Germans about type 2 diabetes in Germany stratified by sex and report the findings in terms of the ratio of false positive diagnoses of type 2 diabetes (FPR) in the data set.

The age-specific FPR for men and women changes with age. In men, the FPR increases linearly from 1 to 3 per mil in the age 30 to 50. For ages between 50 to 80 years, FPR remains below 4 per mil.

After 80 years of age, we have an increase to about 5 per mil. In women, we find a steep increase from age 30 to 60, the peak FPR is reached at about 12 per mil between 60 and 70 years of age. After age 70, the FPR of women drops tremendously. In all age-groups, the FPR is higher in women than in men. In terms of absolute numbers, we find that there are 217 thousand people with a false-positive diagnosis in the data set (95% confidence interval, CI: 204 to 229), the vast majority women (172 thousand, 95% CI: 162 to 180).

Our work indicates that possible false positive (and negative) diagnoses should appropriately be dealt with in claims data, e.g., by inclusion of age- and sex-specific error terms in statistical models, to avoid potentially biased or wrong conclusions.

INTRODUCTION

Aggregated data about the prevalence and incidence of chronic conditions become more and more available for research purposes. Usually, such data refer to a survey period and are presented in age- and sex strata. A prominent example is the National Health And Nutrition Examination Survey (NHANES) conducted by the National Center for Health Statistics at the Centers for Disease Control and Prevention [CDC 2021]. For public health research, NHANES surveys health and nutritional data from the US general population since 1971. Another example is the Global Health Data Exchange (GHDx) catalog comprising three decades of data about the most prevalent and severe diseases from all over the world [GHD 2021]. Regional databases may contain health data from millions of people. In Germany, for instance, it is planned that aggregated claims data including medical diagnoses from the statutory health insurance with roughly 70 million insurants will be published on a regular basis [FDZ 2022]. Given the large number of study participants at possibly many points in time, validity of the diagnoses in such big data sets can hardly be assessed. By validity of diagnoses, we refer to two types of error that may occur: On the one hand, people with the chronic condition in reality might not have the diagnosis coded in the data set and can be assumed to be false negatively coded. On the other hand, people without the chronic condition in reality might have a corresponding diagnosis in the data set. Henceforth, we refer these as false positive findings in the data set. By opposing the diagnoses coded in the data set with "reality", i.e. the "gold standard", such as a medical diagnosis based on an extensive examination of a specialist, the diagnosis codes in the data set can be interpreted similarly as a diagnostic test. Table 1 shows the possible combinations of disease status according to the gold standard and a coded diagnosis in the data set.

| Claims data | Gold standard | |
|---|---|---|
| | Diseased | Not diseased |
| Diagnosed | True positive | False positive |
| Not diagnosed | False negative | True negative |

**Table 1: Possible combinations of disease status and coded diagnoses in the data set.**

Interpreting the diagnoses in the data set as the result of a diagnostic test compared to the gold standard, leads to the concept of diagnostic accuracy, i.e., sensitivity and specificity for the aggregated data. As in diagnostic tests, the percentage of false positive diagnoses in the data set, i.e., the false positive ratio (FPR) and the specificity add up to 100%. The same holds true for the false negative ratio (FNR) and the sensitivity.

In this study, we use a recently proposed method to estimate the FPR in the context of aggregated data about the prevalence, incidence and mortality. The core idea is to relate the temporal change of the prevalence with the incidence and the mortality information by a partial differential equation (PDE) [Brinks & Landwehr 2014]. In order to make the PDE consistent with the empirically observed prevalence, incidence and mortality data, FPR and FNR of the data are needed [Brinks et al. 2021]. With the assumption that the FPR and FNR in the prevalence and incidence data are the same, we can estimate the FPR in a claims data set comprising type 2 diabetes status in 70 million Germans (85% of the overall population)

METHODS

Before we describe how to estimate the FPR in the claims data, we briefly introduce the methodological approach. Based on the illness-death model for chronic conditions, in [Brinks & Landwehr 2014] we could derive a PDE that relates the temporal change of the age-specific prevalence $p = p(t, a)$, i.e. the proportion of people aged $a$ at calendar time $t$ with the chronic condition, with the incidence rate $i(t, a)$, general mortality $m(t, a)$ and the mortality rate ratio $R = R(t, a)$.

$$(\partial_t + \partial_a)\, p = (1 - p)\, \{i - m \times p\, (R - 1) / [1 + p\, (R - 1)]\} \qquad (1)$$

The mortality rate ratio $R$ is the quotient of the mortality rates $m_1(t, a)$ and $m_0(t, a)$ of people with and without the chronic condition, respectively, i.e., $R(t, a) = m_1(t, a)/m_0(t, a)$. Equation (1) holds true for the true prevalence $p$ and incidence rate $i$. If we assume an observed prevalence $p^{(obs)}$ and an observed incidence $i^{(obs)}$ in the data set (possibly imperfect with respect to diagnostic accuracy), the true prevalence and incidence can be obtained from Equations (2a) and (2b) using the sensitivity (*se*) and specificity (*sp*).

$$p = (p^{(obs)} - 1 + sp_p)/(se_p + sp_p - 1) \qquad (2a)$$

and

$$i = (i^{(obs)} - 1 + sp_i)/(se_i + sp_i - 1). \qquad (2b)$$

In Equations (2a) and (2b), sensitivity (*se*) and specificity (*sp*) of the age-specific prevalence and incidence (indicated by the sub-index) need not necessarily be the same. In data sets where prevalence and incidence stem from different sources e.g. different samples or surveys, the distinction might still be useful. Here we assume that the data generating process of prevalence and incidence are the same, such that we can skip this distinction and assume $se_p = se_i$ and $sp_p = sp_i$ for all ages *a*.

Given the observed prevalence $p^{(obs)}$, observed incidence $i^{(obs)}$, general mortality *m*, mortality rate ratio *R*, and sensitivity $se = se_p = se_i$, we can insert Equations (2a) and (2b) into Equation (1) to estimate the specificity $sp = sp_p = sp_i$ [Brinks et al 2020]. Thus, for known sensitivity *se*, we can calculate $sp = 1 - FPR$ from these data by a functional relation $\Phi$:

$$sp = \Phi(se, p^{(obs)}, i^{(obs)}, m, R). \qquad (3)$$

The exact formula for the functional relation $\Phi$ between *sp* on the left hand side and *se*, $p^{(obs)}$, $i^{(obs)}$, *m*, and *R* on the right hand side of Equation (3), is lengthy and presented together with its derivation in the supplement of [Brinks et al 2020].

Usually, we do not know the sensitivity *se* of the diagnoses in the data set. To overcome this problem, we use a probabilistic approach as in [Brinks et al 2020] and randomly sample *se* from epidemiologically reasonable ranges between 50% and 99.9%. This does not impose a problem, because the functional relation (3) is robust with respect to *se*, which has been demonstrated in [Brinks et al 2020]. We examine how the estimated specificity *sp* changes and present the result as false positive ratio $FPR = 1 - sp$. The FPR is allowed to vary over age, independently for men and women in relevant age range 25 to 85 years. The algorithm requires that the age resolution, i.e., the difference between two consecutive age groups, is coarser than the temporal distance between the two prevalence surveys.

The algorithm described above is applied to the claims data about type 2 diabetes presented in [Goffrier et al 2017]. The claims data comprises about 70 million people during the period from 2009 to 2015. The number of people with a diagnosed type 2 diabetes are 5.8 and 6.1 million in 2009 and 2015, respectively. Prevalence of type 2 diabetes in men and women in these years are reported in 17 age groups (<15, 15-19, 20-24, ..., 80-84, 85-89, 90+). Incidence rates for men and women are reported for the years 2012, 2013 and 2014 aggregated in five age groups (<20, 20-39, 40-59, 60-79, 80+). In a first step, reported prevalence $p^{(obs)}$ and incidence $i^{(obs)}$ are transformed by applying the *logit* function and the natural logarithm (*log*), respectively. Then, the transformed values are fit by the least squares method using a natural spline (*ns*) for age *a* with interaction terms in time *t* and sex *s*, i.e., $y \sim ns(a)*t*s$ where *y* refers to $logit(p^{(obs)})$ and $log(i^{(obs)})$, respectively. Note that we only have aggregated data, which prohibits more sophisticated statistical methods such as negative binomial regression etc. The degrees of the natural splines for the transformed reported prevalence and incidence are determined based on the number of available data points and visual comparison of the fitted functions with the reported input prevalence and incidence data.

For applying the functional Φ as in Equation (3) the general mortality *m* and the mortality rate ratio *R* are required. The general mortality is taken from the Human Mortality Database [HMD 2021]. The mortality rates of men and women in Germany during the five years period 2010-4 are fit by a polynomial of degree two in age *a* to the logarithmized mortality rates in the age range 15 to 95. Impact of sex *s* was implemented by an interaction term, i.e., the model equation reads $log(m) \sim (a^2 + a)*s$. The degree of the polynomial was chosen by visual inspection of the fitted function with the input mortality rates. The age-specific mortality rate ratios *R* for men and women refer to the year 2014 and stem from [Nation Diab Surveillance]. After application of a log-transformation, a natural spline in age *a* has been fit to *R*. Sex *s* is taken into account by an interaction term. Thus, the model reads *log(R) ~ ns(a)\*s*. The degree of the natural spline is again determined based on the number of available data points and visual comparison of the fitted functions with the reported mortality rate ratios *R*.

After these data input and fitting routines, Equation (3) is applied and the associated age-specific FPR for men and women are calculated. Since the prevalence data are given in 2009 and 2015, the temporal difference is 2015 – 2009 = 6 years, and estimates for ages more than 6 years apart are possible. We choose ages to be *a* = 25, 32.5, 40, ..., 77.5, 85.

In order to estimate the absolute number of people with a false-positive diagnosis of type 2 diabetes, we interpolate the *FPR*, the corrected prevalence *p* (according to Equation (2a)) and the number of people *N* in the claims data with their age-distribution to all ages from 20 to 100. Then, the number of people $N^{(fp)}$ with a false positive diagnosis is calculated by

$$N^{(fp)} = \sum_{a=20}^{100} S(a) \times FPR(a), \qquad (4)$$

where *S(a)* is the estimated number of people aged *a* without type 2 diabetes *S* = (1 – *p*) × *N*.

Since we sample 100'000 sensitivity values, we obtain a large number of estimates for FPR in men and women using Equation (3). Accordingly, Equation (4) provides a random distribution of possible values in men, women and total. Empirical quantiles (2.5, 50, and 97.5%) for the resulting distributions are reported.

All calculations are performed in the free statistical software R, version 4.1.0 (The R Foundation for Statistical Computing). The source code and data for running the analysis has been published in the open access repository Zenodo with digital object identifier (DOI) 10.5281/zenodo.5906275. The data from the Human Mortality Database are available after registration only [HMD 2021]. We respect this policy and do not upload the raw mortality data in the Zenodo repository. Instead, in the uploaded source code we present the fitted coefficients of the regression model for the mortality rates. Using the coefficients instead of the raw data, which the coefficients were derived from, guarantees that the code is fully functional without unveiling data protected under a policy. Of course, using the coefficients from the regression model does not affect any of the conclusions drawn in this work, because the results are identical.

RESULTS

The data points in Figures 1 and 2 show the reported prevalence $p$ and incidence $i$, respectively, from the claims data, separately for men (left panel) and women (right panel) [Goffrier et al.] These are opposed to the fitted curves (lines) after applying the logit and log transform to the data points. Similarly, in Figures 3 and 4 the reported mortality rate ratios $R$ and general mortality $m$, respectively, for men (left panel) and women (right panel) are shown together with their fitted curves (lines).

After fitting the input data, i.e., prevalence, incidence, mortality rate ratios and general mortality, we have all data at hand to estimate the age-specific FPR for men and women. For both sexes, 100'000 random samples of the sensitivity $se$ are drawn uniformly from the range 50% and 99.9% and the associated FPR are calculated by Equation (3). The results are shown in Figure 5. Each of the 100'000 age-specific FPR for men (left panel) and women (right panel) are depicted as a line, which at higher ages yield the impression of forming an area of possible values, blue and red, for men and women, respectively. In men, the FPR is less than 6 per mil for all ages. In age groups below 50, the FPR in men increases linearly to about 2.5 per mil. At ages greater than 50 the maximum possible FPR is plateauing with a slight dip at age 77.5 followed by an increase to about 6 per mil. In women, the age specific FPR steeply increases until age 60 and peaks at about 12 per mil. For ages greater than 60, the FPR of women is decreasing again. In all age-groups, the FPR is higher in women than in men.

In terms of absolute numbers of false diagnoses in men and women, we obtain the empirical distributions shown in Figure 5. The median and 95% confidence bounds are reported in Table 2.

|        | Number of patients with false diagnoses of type 2 diabetes (in thousands) | |
|--------|--------|--------|
|        | Median | 95% confidence interval |
| Men    | 39.9   | 31.6 to 47.3 |
| Women  | 172    | 162 to 180 |
| Total  | 217    | 204 to 229 |

**Table 2: Number of patients with falsely diagnosed type 2 diabetes in the claims data.**

Obviously, the vast majority of people wrongly diagnosed with type 2 diabetes in the claims data are women.

DISCUSSION

In this work, we estimate the age-specific FPR of type 2 diabetes in German women and men in a huge claims data base. We use a mathematical relation between prevalence, incidence and mortality for chronic conditions. In order to balance this relation, false positive and false negative diagnoses in the claims data need to be considered. Usually, the amount of false positive and false negative diagnoses is not accessible in such data. However, as the false positive findings dominate the impact of false negative findings by magnitudes, making coarse assumptions about the percentage of false negative findings allows to examine FPR.

We find that the age-specific FPR of men and women differ substantially. Across all age groups, FPR in men is lower than in women at the same age. At 60 years of age, FPR of women is at least 3 times as high as the FPR of men. As a consequence, across all age-groups about 172 thousand women and 40 thousand men have a false-positive diagnosis in the claims data. Reasons for the differences can only be speculated about. For example, in Germany women are known to visit a physician more frequently than men [Hoebel et al 2016]. It seems plausible that less frequent contacts decrease the probability of making a false positive diagnose in the claims data. Reports about false positive diagnoses of type 2 diabetes in huge data bases are rare. More than a decade ago, a project about quality improvement for diagnoses of type 2 diabetes using computerized algorithms, found a percentage of false positive diagnoses of about 5% in primary care patient records [de Lusignan et al 2010], which is similar to the percentage of 3.7% found here (217000/5.8 mio = 0.037). The authors conclude their report with the note that the current practice of coding diabetic diagnostic data probably overestimates the prevalence of diabetes. We come to the same conclusion in the claims data examined here, but should remind ourselves to the huge number of people with undiagnosed diabetes in Germany. In a representative population survey 2008-11, the prevalence of diagnosed and undiagnosed diabetes in the overall population has been estimated to be 7.2 and 2.0 percent, respectively [Heidemann et al 2015]. Applying these findings to the claims data, we (roughly) estimate a number of 70 million × (100% – 7.2%) × 2.0% = 1.3 million people with a missing diagnosis of type 2 diabetes in the claims data. Compared to the estimated 190 thousand people with a false positive diagnosis, false negative diagnoses (undiagnosed) are a greater problem in the claims data than false positive findings. Unfortunately, we have not found any report about the differences of false positive diagnoses of type 2 diabetes between men and women.

The main purpose of this work addresses the question how (large) secondary data can be used for epidemiological analyses. Frequently, claims data are easily accessible for large populations. Thus, drawing conclusions relies on large numbers of cases, which seemingly provides an enormous statistical power and a large potential for scientific analyses. However, it is clear that claims data and diagnoses within these data are collected for non-scientific purposes such as documentation and reimbursement. Although making faulty diagnoses, e.g., by coding a tentative diagnosis as an ascertained one, is considered as fraud by national law, it is clear that scientific quality criteria are rarely met. In the claims data about type 2 diabetes, there is no difference between diagnoses of general practitioners and specialists. Validation studies on individual patient level come to the conclusion that diagnoses in claims data have contextual problems, which requires careful and critical analysis [Horenkamp-Sonntag 2017]. Our analysis provides insights into diagnostic accuracy, especially into the amount of false positive diagnoses of type 2 diabetes.

Our method has several advantages. First, the approach described here can be applied to other chronic diseases and requires aggregated data only. Hence, the method may be used in a variety of settings where individual data are hard to obtain, for example, when strict data protection rules apply. Second, the method is flexible about the data sources. Data about the general mortality may frequently be obtained from the national statistical offices. If, furthermore, only prevalence and incidence data are available, the missing data about the mortality rate ratio might be taken from comparable populations, where it is available. An old argument states that the mortality rate ratio provides a stable measure in a wide variety of human populations [Breslow & Day 1980].

Another advantage of the analysis presented here may be seen in the fact that the data used refer to the same population, i.e., insurants of the statutory health insurance in Germany, and to a similar period (2009-15). Using them in the same analysis seems reasonable.

The approach described and used has several limitations. Irrespective if prevalence or incidence data are considered, sensitivity and specificity are assumed to be the same for both types of data ($se_p = se_i$ and $sp_p = sp_i$ for all age groups). A justification for this assumption in the diabetes data analyzed here can be seen in the same origin of the underlying diagnoses that have been used to estimate prevalence and incidence. However, the case definitions for a prevalent and incident case differ slightly. In short, a prevalent case is defined as someone having two ascertained diagnoses of type 2 diabetes in the study year 2009 or 2015 [Goffrier et al]. An incident case has been defined as someone who has two diagnoses of type 2 within a year during 2012-14 but is without diagnosis in the three preceding years. It is not guaranteed that these definitions are consistent in all aspects. For example, it might happen that someone registered as incident case in 2014 might not be counted as prevalent case in 2015. Here, we make the implicit, but untested, assumption that these cases are rare. In theory, the assumption of same sensitivity and specificity in both types of data can be released by applying Equations (2a) and (2b) with $se_p \neq se_i$ and $sp_p \neq sp_i$.

Another limitation comes from the fact that the data used to estimate the mortality rate ratio [Nat Diab Surv] is not exactly the same as the data used for prevalence and incidence [Goffrier]. Although they refer to the same population (people covered by the German statutory health insurance system), the mortality rate ratio ($R$) is estimated on inpatient and outpatient diagnoses while prevalence and incidence refer to outpatient diagnoses only. One might think that for type 2 diabetes the differences are small, but strictly speaking this has not been validated. Moreover, estimation of the mortality rate ratio has been accomplished irrespective of the problem of false-positive and false-negative diagnoses in that data set. Thus, we implicitly assume that the estimates of the age-specific mortality rate ratios are not affected by imperfect diagnostic accuracy. Until now, systematic examination of the quality of mortality estimates from these claims data are missing. A last drawback should be mentioned: prevalence, incidence and mortality rate ratio are estimated on the roughly 70 million people within in the statutory health insurance. The general mortality, however, refers to the overall population of Germany (82 million people). Recent analyses indicate that in ages below 90 years there are no differences between the age-specific mortalities between these groups, see Figure 2 in [Schmidt et al 2021].

To sum up, we underpin the importance of considering false positive and false negative findings in secondary health data. As an application of our considerations, we obtain an estimate of the number of false positive diagnoses in claims data covering about 85% of the German population.

FIGURES

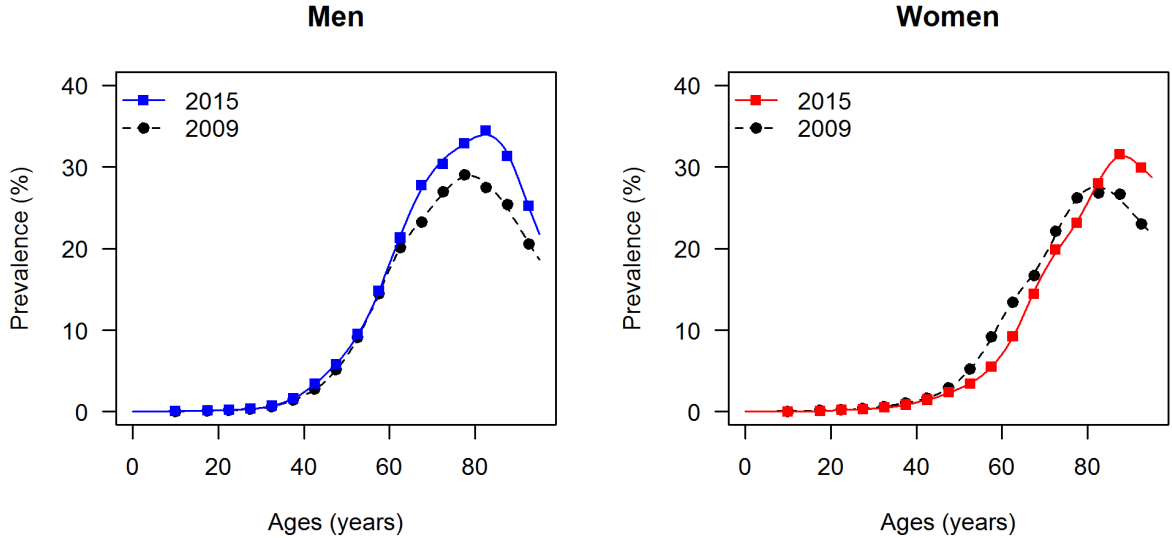

Figure 1: Age-specific prevalence of type 2 diabetes for men (left panel) and women (right panel) in the years 2009 and 2015. The data points and curves are the reported values from the claims data [Goffrier et al] and the fitted functions (lines), respectively.

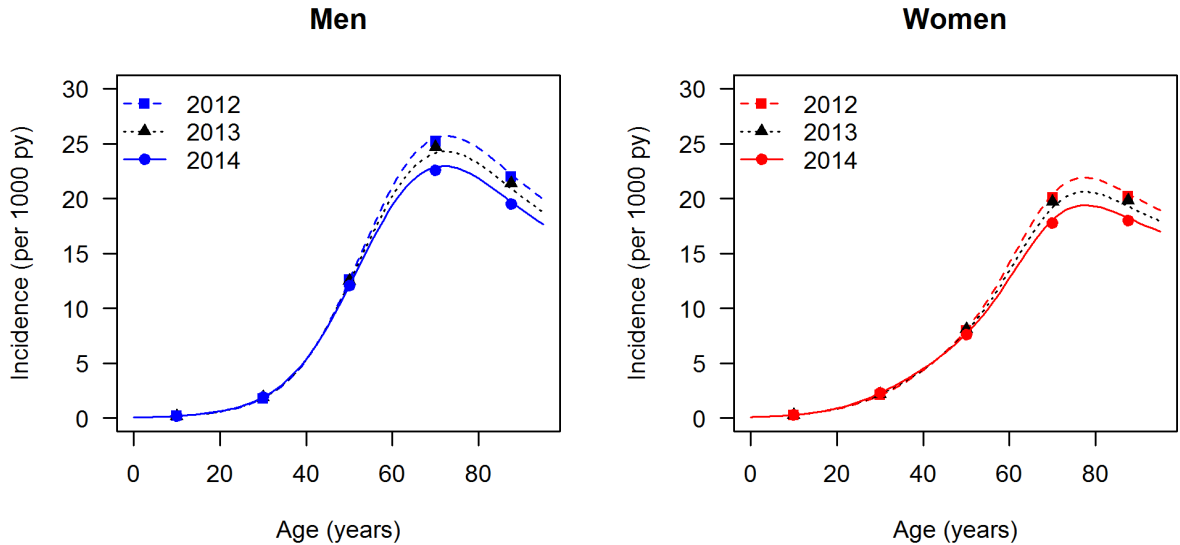

Figure 2: Age-specific incidence of type 2 diabetes for men (left panel) and women (right panel) in the years from 2012 to 2014. The data points and curves are the reported values from the claims data [Goffrier et al] and the fitted functions (lines), respectively.

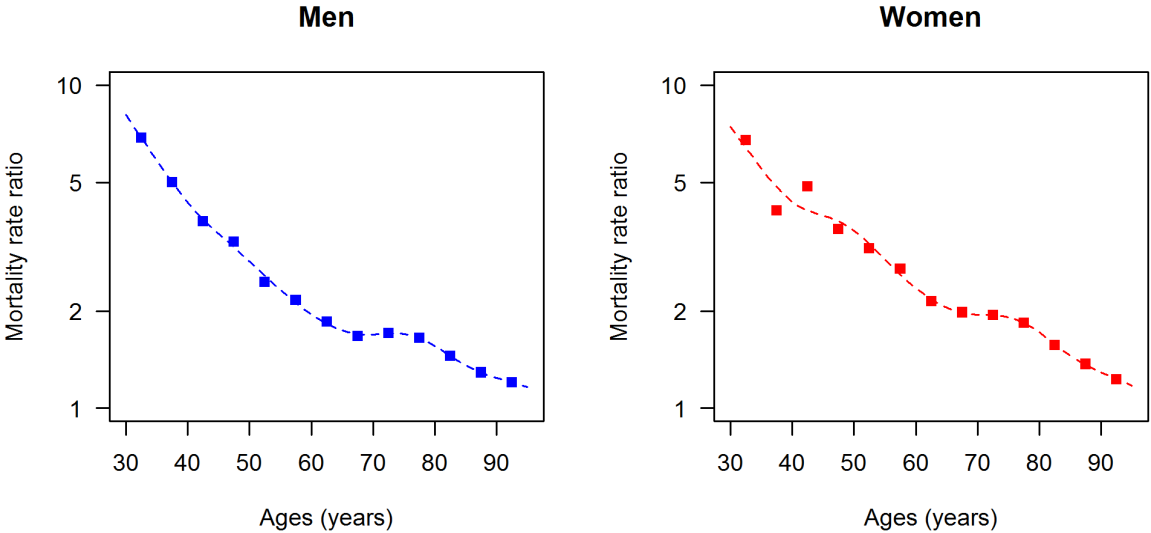

Figure 3: Age-specific mortality rate ratio (diabetes over non-diabetes) for men (left panel) and women (right panel) in the years from 2012 to 2014. The data points and curves are the reported values from the claims data [Nationale Diabetes Surveillance] and the fitted functions (lines), respectively.

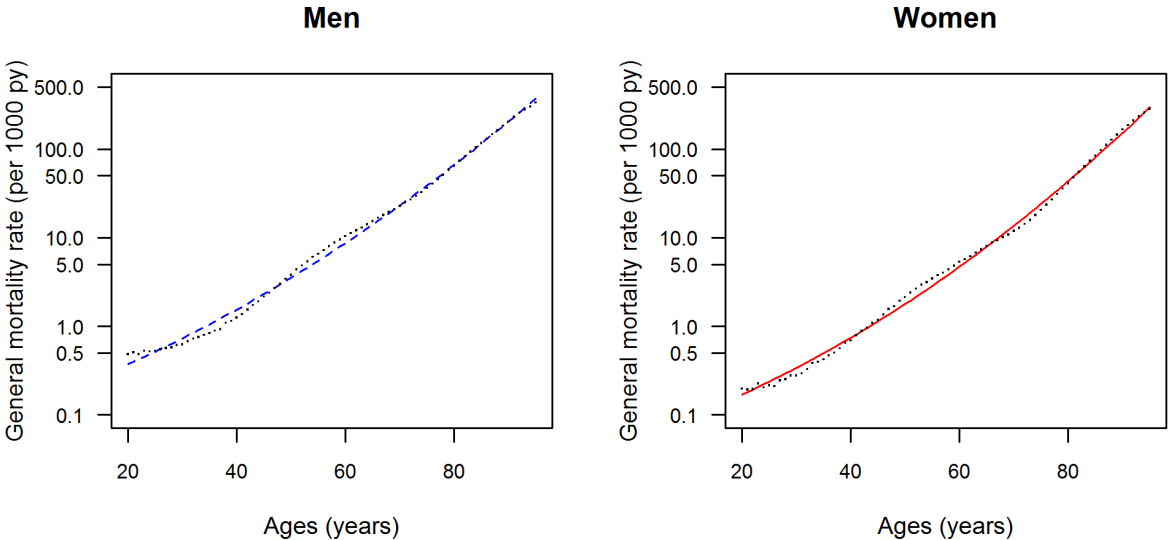

Figure 4: Age-specific general mortality rate for men (left panel) and women (right panel) in the years

from 2010 to 2014. The data points and curves are the reported values from the Human Mortality Database [HMD 2021] and the fitted functions (lines), respectively.

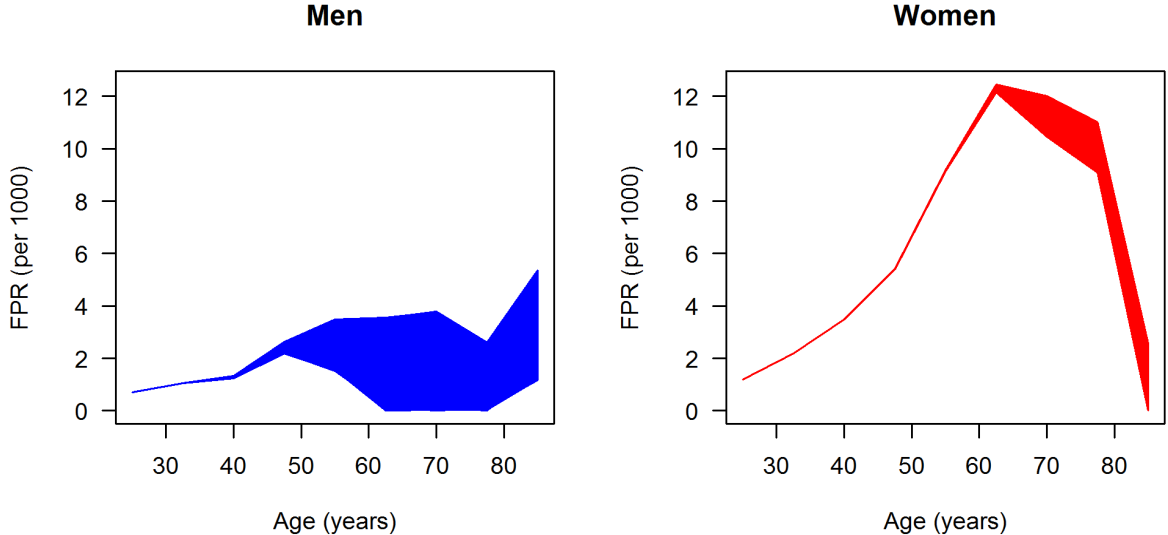

Figure 5: Age specific ratio of false positive diagnose (FPR) in men (left panel) and women (right panel). 100000 (random) scenarios about the age-specific sensitivity are simulated, each line represents the FPR generated by one of these scenarios.

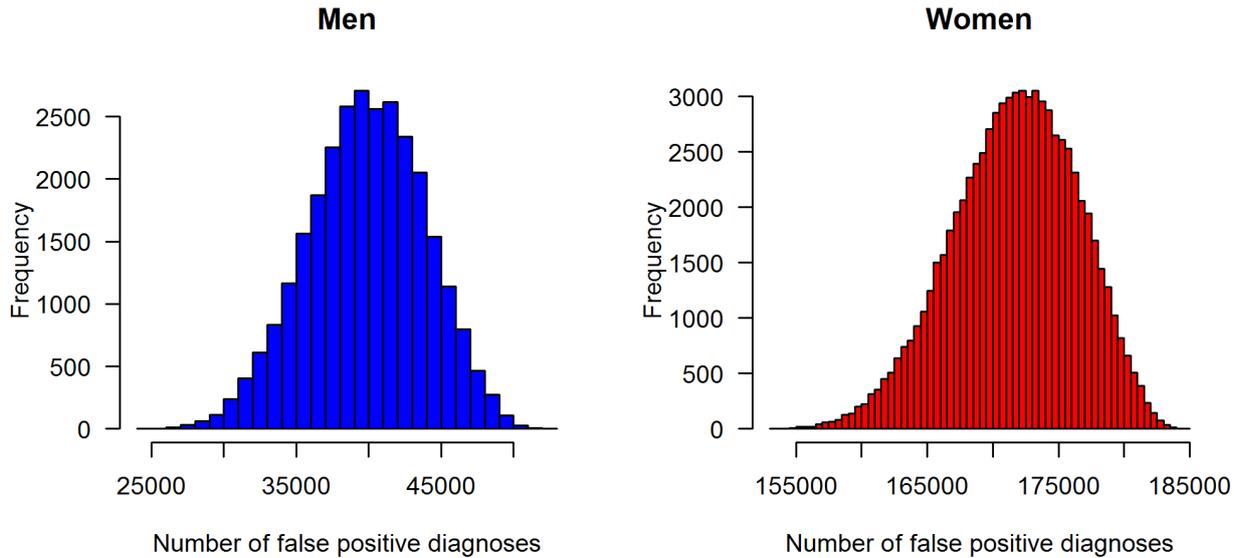

Figure 6: Distributions of the number of false positive diagnoses of type 2 diabetes in the claims data stratified by sex: men (left panel) and women (right panel).